\documentstyle[prl,aps]{revtex}

\begin{document}
\draft
\twocolumn

\input epsf
\title{Inference and Chaos by a
Network of Non Monotonic Neurons}

\author{David R.C. Dominguez\cite{email}}
\address{Theoretical Physics Department C-XI,
Universidad Auton\'oma de Madrid\\
Cantoblanco, 28049 Madrid, Spain}

\date{\today}
\maketitle

%\hfill {preprint FTUAM95/40}

\begin{abstract}

The generalization properties of an attractive 
network of non monotonic neurons which infers 
concepts from samples are studied.
The macroscopic dynamics for the overlap 
between the state of the neurons with the 
concepts, well as the activity of the neurons, are 
obtained and searched for 
through its numerical behavior.
Complex behavior leading from fixed points to chaos 
through a cascade of bifurcation are found, 
when we increase the correlation between 
samples or decrease the activity of the 
samples and the load of concepts,
or tune the threshold of fatigue of the neurons.
Both the information dimension and the
Liapunov exponent are given,
and a phase diagram is built.

\end{abstract}


\label{sec:level1}

\section{Introduction}

There are two sources for building more sophisticated
models of brain behavior as associative memory
other than the original Hopfield model for 
neural networks.
One is the closeness to realistic facts
observed in neural systems,
another one is the trial to attain more complex
learning abilities.
Among the successful attempts for the former
are the multi-state neuron 
models\cite{Ye89}, 
which include
three-state, analog and non-monotonic neurons.
The capability of generalization, the inference of rules
from examples, is a instance of the 
latter\cite{WR93},\cite{OH95}.
The categorization, or capability to retrieve
patterns of activity in different levels
of an hierarchical classification 
is another instance\cite{FM89}.
Here, we work out a connection between 
the multi-state neural networks and
the categorization networks, 
which leads to a new kind of generalization, 
as a property of such neural devices to infer a 
full concept from small samples of that concept.
While in most the neural models of learning 
(see ref.\cite{SS92} and references therein),
the generalization function measures the ability
of the network to give right answers to each question, 
after being trained with samples of 
question-answer pairs,
in the present model the samples are patterns
which carry information about the concepts, 
which can be identified with the answers.

The multi-state neuron model was introduced to 
account for some degrees of ignorance
of pieces of the full pattern.
It differs qualitatively from the two-state model 
because, in absence of part of the information, 
fewer bits are required to represent the 
$small$ $pattern$, so called, as one picked
up information from 
the active sites, keeping the inactive sites off.
Several models of multi-state neurons
were studied with the Hebbian learning algorithm.
The behavior of the analogue neural network 
was studied first in the case of binary 
memorized patterns\cite{MW90}, and yields
a phase-diagram similar to that of stochastic 
binary 
neurons, replacing the temperature $T$ for
the inverse of the gain parameter (the slope
at origin of the transfer function).
The three-state neural network in the presence 
of three-state uncorrelated patterns was studied
within the extremely diluted synapse scheme,
showing an enhancement of the storage capacity 
with an adequate control of its firing threshold. 
This is more notable when the pattern $activity$ 
(the rate of non vanishing states per sites
of the pattern) is small\cite{BV93}.
Non-monotonic neural networks, which take account
of the fatigue of each neuron after being exposed 
to an large post-synaptic potential, was studied
by means of a signal-noise analysis\cite{SF93}.
This network exhibits an interesting 
super-retrieval phase, with vanishing error even
for extensive number of learned patterns.
If it is allowable to the neurons decide exchange 
its states by the opposite of the signal of 
its local field,
the capacity of the network becomes even larger 
than that of three-state neurons\cite{Ko91}.

For all these cases a parallel deterministic dynamics was
assumed given 
by the set of equations

\begin{equation}
\sigma_{it+1}=F_{\theta}(h_{it}),\,\,i=1,...,N
\label{1.1}
\end{equation}
where $\sigma_{it}$\, is the neuron state of site 
$i$ at time $t$, $\theta$ is the threshold parameter 
which represents deviation of the signal function,
and as usual, only odd bounded I/O (Input/Output)
$F_{\theta}$ functions are  considered. 
The local field of site $i$ at time $t$ is

\begin{equation}
h_{it}=\sum_{i(\neq j)}^{N}J_{ij}\sigma_{jt},
\label{1.2}
\end{equation}
$J_{ij}$ being the elements of the synaptic matrix.
In the case of three states neurons and patterns,
the existence of a threshold for which the retrieval
is optimized was also found by 
statistical mechanical techniques within 
the replica symmetric approximation\cite{BR94},
but it can not be useful for the non-monotonic
network, since it does not have an energy 
function\cite{Op95}.

The task of generalization by a neural network can 
be realized in a manifold of contexts. 
One kind is the categorization, which takes place 
if we use an alternative Hebbian learning 
algorithm which stores $s$ examples having 
correlation $b$ with one hierarchical ancestor, 
for each of the $p$ concepts. 
For the connected model, in the context of an
attractor neural network, the following modified 
Hebbian learning algorithm has been 
studied\cite{FM89}:

\begin{eqnarray}
 J_{ij}={1\over N}
\sum_{\mu}^{p}\sum_{\rho}^{s}
\eta^{\mu\rho}_{ i}\eta^{\mu\rho}_{ j}.
\label{1.3}
\end{eqnarray}
The correlation of the learning example 
$\eta^{\mu\rho}$ with one concept of the
set $\{\xi^{\mu}\}$ is 
$<\eta^{\mu\rho}_{i}\xi^{\nu}_{j}>
= b\delta_{\mu\nu}\delta_{ij}$.
The phase transition from an disordered
to a generalization phase, 
where the neurons retrieve one concept, 
was found to be discontinuous with $b$
for a fully connected network\cite{Mi91}, 
or smooth for a diluted network\cite{CT95}. 
After sufficiently increasing $s$ or $b$, 
and decreasing $\alpha\equiv p/N$, 
the error in the generalization became small 
enough to consider such task successfully.

Another interesting kind of generalization 
is inference. 
The coherence between the learned patterns  
with activity $a\ll 1$ allows many patterns 
being simultaneously retrieved\cite{MH89}.
Then, by learning small patterns, we can infer 
the existence of a whole pattern, with activity 
$a\sim 1$. 
Enlarging the effective size of the pattern, 
we can extract much more information than 
the original patterns contain.
For instance, we would see wood where before we 
had only seen trees.
To obtain such an inferential property, however, 
a more sophisticated algorithm is required.
Fortunately, it comes from a modified version 
of the Hebbian algorithm in Eq.(\ref{1.3}).
Nevertheless, it requires a
mathematically difficult effort to 
make a connection between generalization and 
multi-state neurons.
The unique investigation treating the 
generalization with analog neurons\cite{ST92}
uses binary examples.
Then it is worth analyzing such models in 
their simpler, extremely diluted version, 
which yields an exactly soluble dynamics
and is biologically relevant at the same 
time\cite{DG87}.
In this version, a network of three-states
monotonic neurons shows a clear improvement
of the performance as a generalization
device, if small activity examples are
learned\cite{DT95}.

We describe the model of a network of 
non-monotonic neurons in the next section.
After obtaining the recursion relations 
for the inferential properties in the section (III), 
in the last section we present our conclusions,
drawing the curves of generalization with 
special attention at the non-steady solutions.

\section{The model}

We adopt the dynamics given in 
Eqs.(\ref{1.1}-\ref{1.2}), 
and start by defining an I/O function.
Although most works employ stair-like 
(modelling by the $q$-Ising network) 
or other monotonic functions 
$F_{\theta}$,
we will to avoid this restriction
and choose instead

\begin{eqnarray}
F_{\theta}(x)\equiv sgn(x)&,&\,|x|<\theta,
\nonumber\\
0&,&\,|x|\geq\theta.
\label{2.1}
\end{eqnarray}
Thus, the I/O function tell us the way in 
which the network 
updates each neuron, which become fatigued
outside of the interval $|h_{it}|<\theta$,
according to Eq.(\ref{1.1}).

For the synaptic interactions we will assume 
the Hebbian algorithm in Eq.(\ref{1.3}), 
but the examples to be learned 
will be three-states variables, 
like the neuron state itself.
In order to preserve the odd symmetry of the neurons, 
those patterns are uniformly
distributed around the zero state.
Thus the examples $\eta^{\mu\rho}_{ i}$
are independent random variables built from the 
$concepts$ $\xi^{\mu}_{i}$ through the 
following stochastic process:

\begin{equation}
\eta^{\mu\rho}_{ i}=
\xi^{\mu}_{i}\lambda^{\mu\rho}_{ i},
\,\,<\lambda^{\mu\rho}_{i}>\equiv b,
\,\,<(\lambda^{\mu\rho}_{i})^{2}>\equiv a,
\label{2.2}
\end{equation}
where $\xi^{\mu}_{i}=^{+}_{-}1$ 
with equal probability.
The new random variables introduced here, 
$\lambda^{\mu\rho}_{ i}$, 
are characterized by their mean $b$
and their square mean $a$,
for all examples $\eta^{\mu\rho}$. 

Then the parameter $a$ is the 
$activity$ of the examples them self,
while $b$ is the $correlation$ 
between examples and their respective concept.
On the one hand we can recover the pure generalization 
model\cite{FM89} by setting 
$\lambda^{\mu\rho}_{i}=^{+}_{-}1$ 
($a=1$) with a bias $b$ for the positive value,
and threshold $\theta\to\infty$. 
In this simple limit the neurons are thought of 
as being  submitted to background noise, 
perhaps due to some dirtiness on the pattern.
On the other hand, the pure multi-state model can 
also be obtained by taking the number of examples 
$s=1$ in Eq.(\ref{1.3}) and correlation $b=1$.
A low activity $a\ll 1$ indicates that in many sites 
the patterns are not active, 
$|\eta^{\mu\rho}_{i}|\neq 1$, 
with the effective size 
of the learned patterns being $N_{e}=aN$. 
So, when the activity $a$ is not close to $1$, we 
can speak of a $small$ $pattern$\cite{Ye89}.
In our model the new viewpoint is the following: 
the small examples are $samples$ of the full activity
concepts to be inferred.

The task of generalization (inference) is successful 
if the distance between the state of the neuron and 
the concept $\xi^{\mu}$, defined as
$
E^{\mu}_{t}\equiv 
{1\over N}\sum_{i}|\xi^{\mu}_{i}-\sigma_{it}|
$
becomes small after some time $t$.
This is the so-called Hamming distance, which in
this context is called the
$generalization$ $error$. 
In order to measure the quality of the retrieval 
of the small patterns\cite{BV93}, 
one needs to consider a Euclidean quadratic 
distance instead of the Hamming distance, but we are 
interested exclusively in the capacity of the network 
to infer a larger concept of full activity from 
the samples, in which case $E^{\mu}_{t}$ suffices.

{\em Remark:}
Since $E^{\mu}$ is $\mu$ dependent it looks like a
training error with respect to just one pattern\cite{SS92}.
However, it is not dependent on the examples 
$\eta^{\mu\rho}$, being indeed a generalization error,
which is $p$-degenerate in the concepts,
and it can be chosen a particular state $\sigma$ 
near to $\xi^{1}$.

The relevant order parameters for the dynamics, 
during some specified time $t$, when the state of network 
is given by $\{\sigma_{it}\}$, are the 
$retrieval$ $overlaps$ 

\begin{equation}
m^{\mu\rho}_{Nt}\equiv
{1\over aN}\sum_{j}^{N} \eta^{\mu\rho}_{ j}\sigma_{jt}
\label{2.5}
\end{equation}
of the $\alpha_{th}$-example of the $\mu_{th}$-concept.
They are normalized parameters within the 
interval $[-1,1]$, 
which attain the extreme value 
$m^{\mu\rho}_{ N}=1$ whenever 
$\eta^{\mu\rho}_{ j}=\sigma_{j}$, by virtue 
of Eq.(\ref{2.2}).
Using this definition, with the synaptic interaction 
in Eq.(\ref{1.3}), 
the local field in Eq.(\ref{1.2}) becomes

\begin{equation}
h_{it}= 
a\sum_{\mu}^{p}\sum_{\rho}^{s}
\eta^{\mu\rho}_{ i}m^{\mu\rho}_{Nt}.
\label{2.6}
\end{equation}
Next we need to analyze the evolution of the $p.s$ 
coupled equations
(\ref{2.5}) instead of the $N$ original Eqs.(\ref{1.1}). 

Because we are interested in the generalizing 
property of our network,
we take an initial configuration whose retrieval 
overlaps are 
only macroscopic of order $O(1)$ for the $s$ examples 
of a given concept, let say the first one,
and symmetric (equal for all $\rho$).
We write 
$m^{1s}_{Nt=1} = \sum_{\rho}^{s} m^{1\rho}_{Nt=1}$
for the $symmetric$ overlap.
In the thermodynamic limit, the retrieval overlaps 
$m^{1\rho}_{Nt=1}$ in Eq.(\ref{2.5}) are  
infinite sums of independent random variables (IRV),
whose fluctuations around its mean value
$<<m^{1\rho}_{Nt=1}>>$ can be neglected.
Then the Law of Large Numbers (LLN) applies to get

\begin{equation}
m_{t=1}\equiv\lim_{N\to\infty}m^{1s}_{Nt=1}=
<<x_{s}F_{\theta}(\Lambda_{t=0})>_{x_{s}}>_{\omega_{0}},
\label{2.7}
\end{equation}
which is $i$-site independent.
Here we have defined the new variable of field
$
\Lambda_{t=0}\equiv\xi^{1}\cdot h_{t=0}=
m_{t=0}sa^{2}x_{s}+\omega_{0},
$
where $m_{t=0}$ is the initial symmetric 
retrieval overlap,
$x_{s}\equiv{1\over as}\sum_{\rho}^{s}\lambda^{1\rho}$,
and $\omega_{0}$ is the noise produced by the 
$p-1$ residual concepts in Eq.(\ref{2.5}).
The averages in the brackets are over 
both $x_{s}$ and $\omega_{0}$ terms in the field.
We have used the odd-property of $F_{\theta}$,
and wrote the argument in $F_{\theta}$ here as a 
sum of two different kind of terms.
The first one favors the ordering in direction 
of the first concept, while the second 
$\omega_{0}$ introduces an additional noise to 
the original mistakes represented for 
those sites where $\lambda^{1\rho}_{i}=-1$.

The most interesting feature for us is the generalizing 
property of our network.
It is characterized by the overlap of the neural 
state with the first concept, 
given in the first time step by

\begin{equation}
M_{t=1} \equiv \lim_{N\to\infty}
{1\over N} \sum_{i}\xi^{1}_{i}\sigma_{it=1}
= <<F_{\theta}(\Lambda_{t=0})>_{x_{s}}>_{\omega_{0}},
\label{2.9}
\end{equation}
which is related to the generalization error 
(the Hamming distance) by $E^{1}_{t=1}=1-M_{t=1}$.
For multi-state neurons it is useful to define 
the $dynamical$ $activity$ order parameter, given in 
the first time step by

\begin{equation}
Q_{t=1} \equiv \lim_{N\to\infty}
{1\over N} \sum_{i} (\sigma_{it=1})^{2} =
<< [F_{\theta}(\Lambda_{t=0})]^{2}> _{x_{s}}>_{\omega_{0}}.
\label{2.10}
\end{equation}
It accounts for the active neurons, and plays a 
similar role as the $spin$ $glass$ parameter of 
the thermodynamic equilibrium approach for 
binary neurons, since it allows one to measure the
degree of order even when there is no retrieval at all
\cite{MP87},\cite{TK90}.
In both the last two equations,
we have used the LLN for a sum of IRV, 
with vanishingly fluctuations, in the thermodynamic limit.

\section{Diluted dynamics}

Although it is easy to solve the single time Eq.(\ref{2.7})
and to obtain the generalization error $E_{t}$, 
the recursion relations for any time $t$ are not 
easily solved.
We then use the extremely diluted synapse approximation, 
for which the first time step gives exact results 
for any number of time steps. 
In this limiting situation the synaptic interactions take 
a vanishing value for almost all pairs of neurons $\{ij\}$, 
and are of the form given in Eq.(\ref{1.3}) only for a small 
fraction $C/N\ll 1$ of them.
The Eqs.(\ref{2.7}-\ref{2.10}) are then reproducible for any $t$, 
with the following simple  distribution ($\doteq$) of the
noise caused by the examples of the $p-1$ residual concepts:
$\omega_{t}\doteq z_{p}\sqrt{\alpha Q_{t}r}$, where
$\alpha=p/C$, $r=s[a^{2}+(s-1)b^{4}]$ and $z_{p}\doteq N(0,1)$ 
is a Gaussian random variable with mean $<z_{p}>=0$ and unit 
variance.
$Q_{t}$ is the $dynamical$ $activity$ at time $t$.

We will also use an approximation for the 
case of many examples ($s>10$):
$x_{s}\doteq {b\over a} +
z_{s}\sqrt{{a-b^{2}\over sa^{2}}}$
with $z_{s}\doteq N(0,1)$ independent of $z_{p}$.
With these remarks, after some algebra with both Gaussian
$z_{s}$ and $z_{p}$ we can write the arbitrary time step 
dynamics for the macroscopic parameters, 
with the I/O function given by Eq.(\ref{2.1}).

The dynamical activity is the following:

\begin{equation}
Q_{t+1}=
{1\over 2}[erf(A_{+})-erf(A_{-})],\,\,
A_{^{+}_{-}}\equiv
{m_{t}sab ^{+}_{-}\theta\over
\sqrt{v_{t}}},
\label{3.1}
\end{equation}
with $v_{t}\equiv
sa^{2}(a-b^{2})(m_{t})^{2}+\alpha rQ_{t}$,
and the symmetric retrieval overlap is 

\begin{equation}
m_{t+1} = {b\over a}M_{t+1} + m_{t}(a-b^{2})C_{t+1}
\label{3.2}
\end{equation}
where we have defined 
$erf(x)\equiv\int_{0}^{x}dy \varphi(y),\,
\varphi(y)\equiv exp(-y^{2}/2)/\sqrt{2\pi}$.
Here,

\begin{equation}
M_{t+1}=
erf({sabm_{t}\over\sqrt{v_{t}}})-
{1\over 2}[erf(A_{+})+erf(A_{-})]
\label{3.3}
\end{equation}
is the overlap of generalization, and

\begin{equation}
C_{t+1}\equiv <F_{\theta}'(\Lambda_{t})>_{z}=
{1\over\sqrt{v_{t}}}
[2\varphi({sabm_{t}\over\sqrt{v_{t}}})
-\varphi(A_{+})-\varphi(A_{-})].
\label{3.4}
\end{equation}
We will make no restrictions about the values 
which the parameters $b$ and $a$ can assume 
within the $(0,1)$ interval,
except that they must satisfy $a\geq b^{2}$ 
(the equality corresponding to constant 
microscopic activities $\lambda\equiv b$).

\section{Attractors and conclusions}

Two fixed-point ordered phases can appear: 
namely, the $Generalization$ phase 
$\{G: M>0,Q>0\}$ and the 
$Self-sustained\,\, activity$ 
$\{S: M=0,Q>0\}$
(or microscopic chaotic\cite{BE93}) phase.
However, the most interesting attractors
are the non-steady macroscopic phases.
Although the Eqs.(\ref{3.1}-\ref{3.4})
are deterministic, averaged over the stochasticity
induced by the extensive load $p=\alpha C$,
some complex behavior remains present 
in the large time dynamics.
It appears a $Doubling$ of period 
generalization phase
$\{D: M_{t}>0,Q_{t}>0\}$, 
without fixed-point, 
where cyclic or chaotic attractors arise.
It can be viewed in the curves of Generalization
showed in the figures below.

In the Fig.1(below) we see the generalization error
$E_{t}$ dependence on the sample correlation $b$, 
and activity $a$, in which we took $a=b$.
Fixed values of the number of examples,
load rate and threshold of fatigue are used
When $b$ is increased until $b_{1}\sim 0.19$, the 
generalization error has a fixed-point behavior.
It initially falls until a optimal  value 
$E_{t}\sim 0.07$ at $b_{op}\sim 0.15$.
Then it reaches a first bifurcation, 
beyond which it oscillates between two values, 
exhibiting a periodic behavior.
A cycle-4 is found after a second bifurcation
at $b_{2}\sim 0.31$, and this doubling of period
follows until a quasi-periodic behavior
takes place at $b_{\infty}\sim 0.35$.
Between $b_{\infty}<b<b_{S}$,
regions of chaos intercalate with
windows of periodicity.
After $b_{S}\sim 0.6$, 
although the correlation is large, 
the activity is large too, 
and it destroys the capacity of generalization,
so that $E_{t}=1$.
The same behavior was qualitatively found
as a function of activity $a$ ($b$), 
keeping fixed $b$ ($a$).
For sufficiently low (high) activity (correlation),
$E_{t}$ oscillates aperiodicaly, 
eventually closer to each chosen 
initial value but never equal to it.

In order to measure the degree of the 
non-regular behavior
we calculated the Liapunov exponent
in the region of $a=b$ above.
It was estimated as\cite{ER85}
$\lambda_{L}\sim
{1\over T}\ln[\delta m_{T}/\delta m_{0}]$,
for $T\gg 1$, where 
$\delta m_{t}$ is the distance between
two trajectories initially near to each other. 
It gives positive values within the
interval $b_{\infty}<b<b_{S}$,
attaining the value $\lambda_{L}\sim 0.34$
at $b_{C}\sim 0.41$
as we can see in the Fig.1(above).
It indicates how chaotic is the oscillation
of $E_{t}$ in this attractor,
which shows sensitivity to initial conditions. 
We also calculated the information 
dimension of the attractor, 
estimated by\cite{ER85}
$d_{H}\sim \ln(N_{r})/|\ln(r)|$, $r\ll 1$,
where $N_{r}$ is the number of balls with
radius $r$ necessary to cover all points $E_{t}$.
For the point $b_{C}$ we got $d_{H}=0.81$.
The non integer value of $d_{H}$ shows that
such attractor is a fractal.

The behavior as a function of $\theta$
is drawn in the Fig.2(below), 
where the effect of the fatigue is singled. 
When the threshold is small enough
the generalization is bad because the
local fields almost everywhere
exceed $\theta$, 
which lead the neurons to its fatigue phase.
After $\theta_{1-}\sim 1.3$ the probability 
of the local field being lower than $\theta$
becomes relevant, then a periodic regime start.
A chaotic regime happens between
$3.8<\theta<6$ when the
local fields fluctuate around $\theta$.
An atypical exit from the chaotic regime
occurs when the $\theta$ is so big that the
local fields gradually leave the non-sigmoidal
phase until at $\theta_{1+}\sim 15$
a new fixed point regime sets in, but now
with a good generalization.

A bifurcation diagram was also found as
a function of the load rate of concepts
$\alpha$.
The noise induced by the saturation of concepts 
rose a large fluctuation for the local fields. 
Thus the chaotic behavior, which implies a 
very sensitive flow of the neural states
with their previous states, is lost for large
$\alpha$.
A phase diagram of the model is shown
in the Fig.2(center), for fixed
values of $a,b,s$.
For small values of $\alpha$, a transition
from a $S$ phase to a $D$ phase occurs, 
whenever the threshold of fatigue crosses 
the solid curve. 
For larger values of $\alpha$, 
the solid curve separates the 
$S$ phase from a $G$ phase.
The $G$ phase is separated from the $D$
phase by the dashed curve.
Differently from the phase diagram
obtained in \cite{BV96},
here no phase $\{Z:M=0,Q=0$\} 
can be reached,
as can be seen from the
Eq.(\ref{3.1}), with $m_{t}=0$, 
which reads
$Q_{t+1}= 
erf({ \theta \over \sqrt{\alpha rQ_{t}} })$.
For $Q_{t}\to 0$ we get $Q_{t+1}\to 1$.
The $S$ phase competes in one region
with the $G$ phase,
but the last is more stable overall 
in this region.

In order to compare with the monotonic case, 
for which the I/O function 
$F_{\theta}(x)\equiv sgn(x),\,|x|>\theta
\,(\equiv 0,\,|x|\leq\theta)$
can be taken, we built
the phase diagram $\alpha(\theta)$
of the Fig.2(above).
The parameter $\theta$ here represents a
threshold of fire of the neurons.
There are no $D$ phase for this case,
but instead, a $Z$ phase can appear for
large enough values of $\theta$.

It is not too surprisingly that the motion 
of the neuron states themselves 
can be over a chaotic trajectory, 
where the memory of the initial configuration 
is not preserved.
But in this case the macroscopic parameter
measuring the retrieval of one pattern is
$M_{t}=0$ almost always, 
because the motion is ergodic over the trajectory, 
running equally over all possible state, 
the huge majority of which have vanishingly 
overlap with that pattern.
This is the case of the $S$ phase.
In the present model, however, the chaos appear
on the less complex macroscopic trajectories for
the overlap in so manner that almost always 
$M_{t}>0$. 
Then we can conjecture that
in the non-steady regimes, the network
preserve a memory of what concept was used
as a seed on the initial configuration.
Thus it can not be related to the properties
of sequential generalization\cite{LC94},
for which a set of
concepts can be retrieved consecutively.
Because the vector of overlaps $\vec{M}_{t}$
can be roughly orthogonal to its previous state,
many other directions $M^{\mu}_{t}$
become macroscopic in each time.
Only one concept, however, is persistently 
retrieved, at varying magnitude.

A similar result was recently found for the
pure multi-state model for retrieval of patterns,
but using analog non-monotonic neurons
instead of our discrete neurons\cite{BV96}.
This shows that the present complex behavior
is rather a consequence of the non-monotonicity
than a characteristic of the generalization model.

The diagrams in Figs.1-2 demonstrate 
how a network of non-monotonic neurons 
can exhibit a complex behavior.
The coherent retrieval of samples leads to the
ability to infer a large activity concept,
even for a large load ratio.
The periodicity of the generalization can be  
controlled by the activity of the samples,
their correlation with each other, 
and the gain parameter of the neurons.
We hope it is worth verifying such behavior
of the inferential properties with 
other learning algorithms and higher levels
of hierarchy.

{\bf Acknowledgments}

This work was financially supported by Cnpq/Brazil.\\

%\newpage

\begin{figure}[t]

\begin{center}
\epsfysize=8cm
\leavevmode
\epsfbox[1 1 600 800]{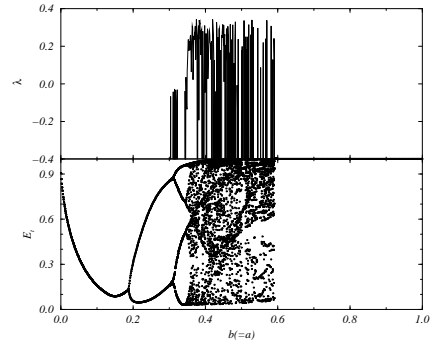}
\end{center}

\begin{figure}
\caption{ {\em Below:}
Generalization error $E_{t}$
as a function of the pattern
correlation and activity $a=b$,
number of examples $s=20$, 
load rate $\alpha=0.01$,  
and threshold of fatigue $\theta=1$.
{\em Above:} 
The Liapunov exponent for
the attractor of the figure below. }
\end{figure}

\label{1EL}

\end{figure}

\begin{figure}[t]

\begin{center}
\epsfysize=8cm
\leavevmode
\epsfbox[1 1 600 800]{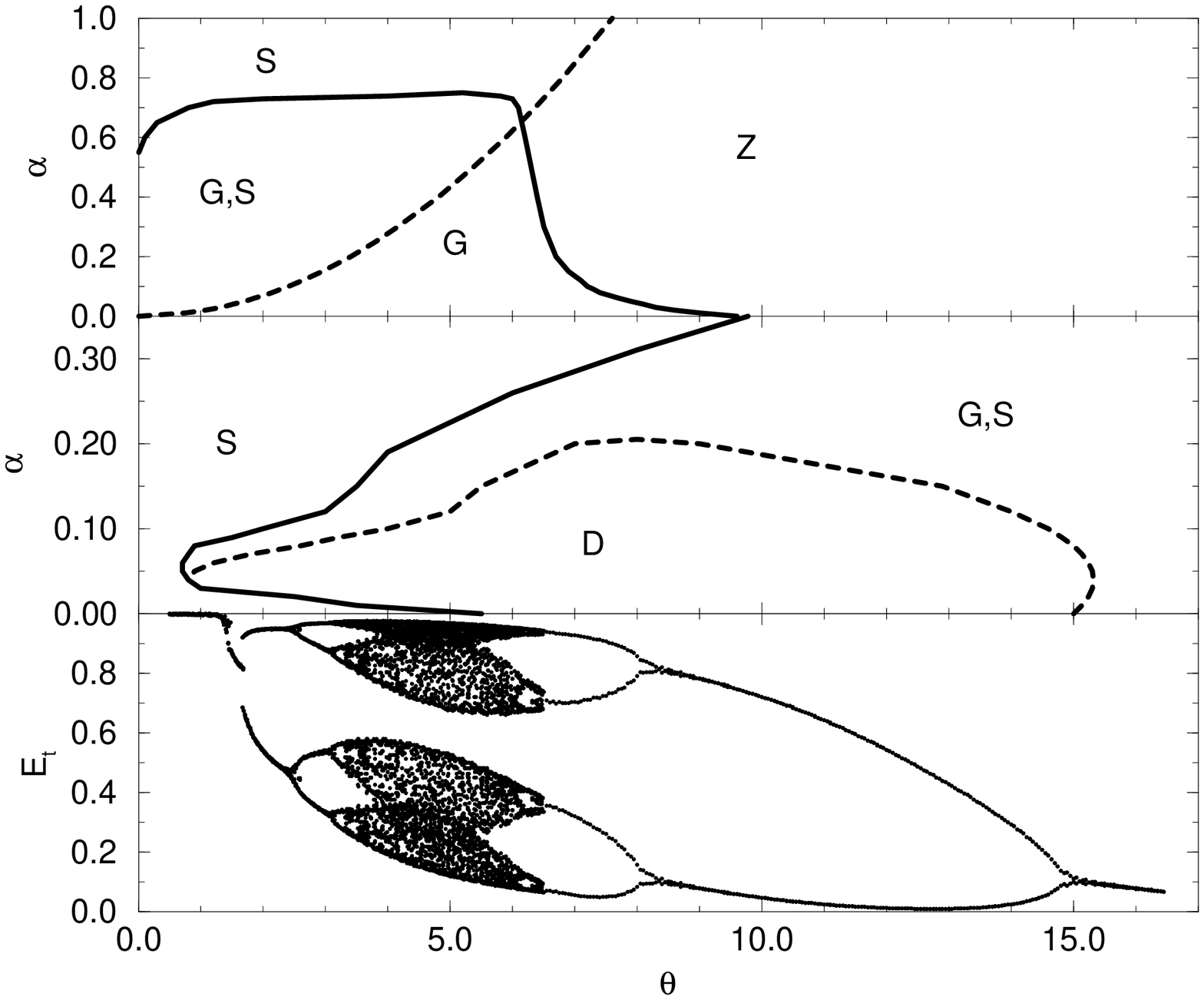}
\end{center}

\caption{ {\em Below:}
Generalization error $E_{t}$
as a function of $\theta$ with $b=0.5=a$,
$s=50$, $\alpha=0.05$.
{\em Center:}
The phase diagram $\alpha(\theta)$, with
$b=0.5=a$ and $s=50$.
The dashed curve separate the
$D$ phase from the $G$ phase, 
while the solid curve separate
the $S$ phase from or the
$G$ or the $D$ phases.
{\em Above:} 
The phase diagram $\alpha(\theta)$, with
the same parameters of the figure at the center, 
but with monotonic of three-state neurons. }

\label{2TDE}

\end{figure}

\end{document}